# Dynamic Weight-Based Collaborative Optimization for Power Grid Voltage Regulation

**Cristian Cortés[1], Hamed Valizadeh Haghi[2], Changfu Li[1] and Jan Kleissl[1,2]**

1　Department of Mechanical and Aerospace Engineering, University of California, San Diego (United States)

2　Center for Energy Research, University of California, San Diego (United States)

## Abstract

Power distribution grids with high PV generation are exposed to voltage disturbances due to the unpredictable nature of renewable resources. Smart PV inverters, if controlled in coordination with each other and continuously adapted to the real-time conditions of the generation and load, can effectively regulate nodal voltages across the feeder. This is a fairly new concept and requires communication and a distributed control logic to realize a fair utilization of reactive power across all PV systems. In this paper, a collaborative reactive power optimization is proposed to minimize voltage deviation under changing feeder conditions. The weight matrix of the collaborative optimization is updated based on the reactive power availability of each PV system, which changes over time depending on the cloud conditions and feeder loading. The proposed updates allow PV systems with higher reactive power availability to help other PV systems regulate their nodal voltage. Proof-of-concept simulations on a modified IEEE 123-node test feeder are performed to show the effectiveness of the proposed method in comparison with four common reactive power control methods.

*Keywords: PV generators, distributed optimization, distribution network, smart inverters*

## 1. Introduction

Increasing presence of renewable energy in the power grid aims at reducing greenhouse gas emissions, among other benefits, like decreasing the fossil fuel dependency as an energy source for a government (Walling et al., 2008). However, the proliferation of variable renewable distributed energy resources (DER), such as solar photovoltaic (PV), introduce power flow changes that cause voltage profile disturbances (Pecenak et al., 2018). This is especially an issue if the nominal peak capacity of DERs gets closer to or surpasses the distribution network's load size. These problems come from the highly variable and intermittent unpredictable nature of renewable energy sources.

In the case of PV systems, smart inverter (SI) control can help mitigate voltage problems by actively utilizing surplus capacity of the inverter. New standards, such as IEEE 1547 (IEEE Standard Association, 2018) and California Rule 21(California Energy Commission and California Public Utilities Commission, 2015), provide guidelines to use the capacity of these devices to inject or absorb reactive power to/from the grid. The real power of the inverter may also be curtailed to provide enough headroom for reactive power utilization under poor voltage scenarios. Real and reactive power control of SI can help regulate the voltage of PV node connecting to the grid, referred to as the point of common coupling (PCC).

As an alternative to the costly traditional network reinforcement approach, real and reactive power control of SIs (agents) have been studied to resolve voltage disturbances caused by fluctuating renewable DERs. There exist different control strategies to use smart PV inverters for voltage regulation. The simplest one considers that every agent solves its own problem using only local information and applying the solution at its PCC, without any coordination with other agents connected to the same power grid. These are known as decentralized control approaches, like droop control (Katirarei et al., 2008; Kashani et al. 2019). These methods are effective only with few PV systems connected to the distribution network. However, if in the same power grid, the installed PV capacity gets close to the peak load connected to it (high PV penetration), some SIs agents can experience non-optimal voltage.

To overcome issues from high PV penetration, other methods exploit coordination among SIs through communication. This type of approach has been proposed in phase 2 of California Rule 21 (CEC & CPUC, 2015). The centralized approach requires a central entity to collect data from all SIs, calculates the optimal operation of each SI (agent), and sends the commands back to each agent (Tsikalakis and Hatziargyriou, 2008; Li et al., 2019). Although the centralized approach requires full communication between the central entity and each agent, coordination of SIs can also be achieved in a distributed manner, which can realize cooperation between different agents without a central entity. For the distributed coordination method, only communication between agents is required, and each agent assigns a parameter (weight) to each link connected to it to define the relevance of one link with respect to the rest. Several works have studied this type of coordination of SIs, considering different communication topologies. A distributed consensus algorithm using adjacent nodes communication is presented in Haque et al. (2019), which determines parameters for the amount of real power curtailed based on the PCC voltage by optimization. In Olivier et al. (2016), a five-step control approach is presented, which combines local regulation with distributed communication for reactive power dispatch and active power curtailment. In those works, fixed communication weights are used, which is not optimal for operation of a dynamic distribution network with variable renewable DERs.

To the authors' best knowledge, this paper is the first one proposing an adaptive reconfiguration of the communication weights for coordinating SIs in a distributed fashion. To decide the fair utilization of each PV unit, the maximum amount of reactive power available each time step is considered. The algorithm utilizes a distributed collaborative optimization to determine optimal reactive power (curtailing real power if it is necessary) and achieves cooperation of SIs for voltage regulation. Effectiveness of the proposed method will be evaluated comparing voltage deviation at each PV system's PCC from simulations. Other currently applied control methods are simulated as well to compare performances. OpenDSS will be used to perform the steady-state simulations of the distribution grid with MATLAB handling the control.

## 2. Problem formulation

To illustrate the problem to be solved in this paper, let's consider a simple case where two PV systems with different sizes (1500 kVA and 75 kVA) are connected to a distribution feeder (Figure 1). A clear day irradiance profile is used for both PVs.

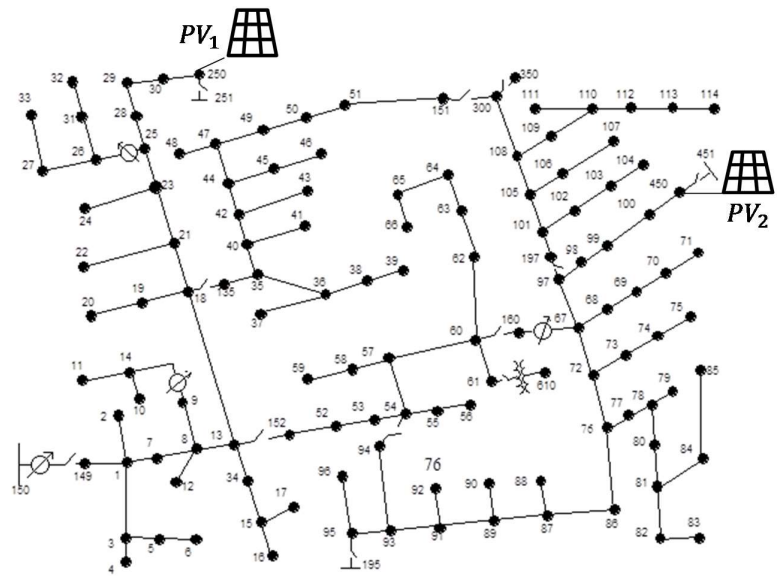

**Figure 1 Two PV systems with smart inverters connected to IEEE 123 node test feeder. 1500 kVA is the rating of the SI rating on $PV_1$ and 75 kVA is the rating of the SI on $PV_2$.**

Without any control (PV operating at unity power factor), the nodal voltage at $PV_1$ goes up to 1.03 p.u. around noon, and the nodal voltage at $PV_2$ is higher than 1.05 p.u. the entire time simulated (Figure 3). SIs can help to regulate the local voltage in this scenario with reactive power. A fixed control method like Volt-Var curve (Figure 2) can be used to achieve this, which establishes a relation between the voltage at the PCC and the amount of reactive power that needs to be dispatch by the SI according to the deviation of the voltage from a reference value ($V_{ref}$). The Volt-Var curve needs parameters as input (IEEE 1457 default setting are used in Figure 2) to define at which voltage level the SI should start dispatching reactive power, and at which rate this is going to increase when voltage deviation rises. $\bar{Q}_m$ is the maximum reactive power available on each SI and is going to be limited by the amount of active power that the inverter is currently injecting, according to the following equation:

$$\bar{Q}_m = \sqrt{S_m^2 - P_m^2}, \qquad \text{(eq. 1)}$$

where $P_m$ is the active power generated by the PV panels connected to the SI $m$, and $S_m$ is the inverter power rating. If more reactive power needs to be dispatched, the inverter can be set in Var-priority mode and active power will be curtailed to use that SI capacity in the additional reactive power needed. To assess the contribution of each SI connected to the distribution feeder, a reactive power fair utilization ratio, $\alpha_m$, is defined as follows:

$$\alpha_m = {Q_m}/{S_m}, \qquad \text{(eq. 2)}$$

where $Q_m$ is the generated reactive power for PV system $m$, and the inequality $Q_m \leq \bar{Q}_m$ always holds.

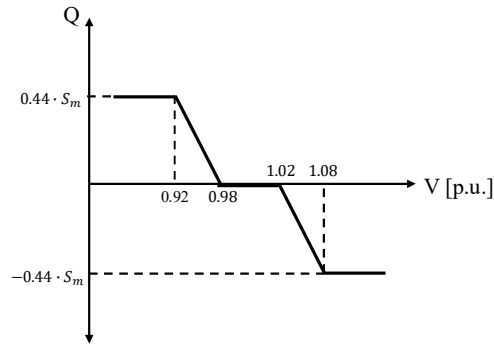

**Figure 2 Volt-Var curve implemented in PV smart inverters to regulate the voltage at PCC using IEEE 1457 default setting. Voltage (V) in per unit, with $V_{ref}$=2.4 kV, and reactive power (Q) to be dispatched on each case.**

If the fixed Volt-Var curve in Figure 2 is implemented, both SIs will absorb reactive power to decrease the voltage at their PCC (Figure 3). Voltage in both PV systems changes a small fraction around noon, and no SI can reach 1 per unit value with this control method (considering $V_{ref}$=2.4 kV). Results for reactive power fair utilization ratio show the percentage of reactive power injected on each case. $PV_2$ is injecting around one third of its SI capacity; meanwhile, $PV_1$ uses only around 10%. Voltage and reactive power profiles for $PV_1$ illustrate that a control method better adjusted to its condition should fix its voltage. On the other hand, $PV_2$ hardly will dispatch enough reactive power to regulate its PCC's voltage down to 1 per unit. Under this scenario, it is valued to propose new methodologies to improve voltage regulation controlling reactive power dispatch at each PV system's PCC. In particular, cooperation between both SIs (using communication capacities) can be applied to manage the extra capacity of $PV_1$ to help voltage regulation at $PV_2$ node.

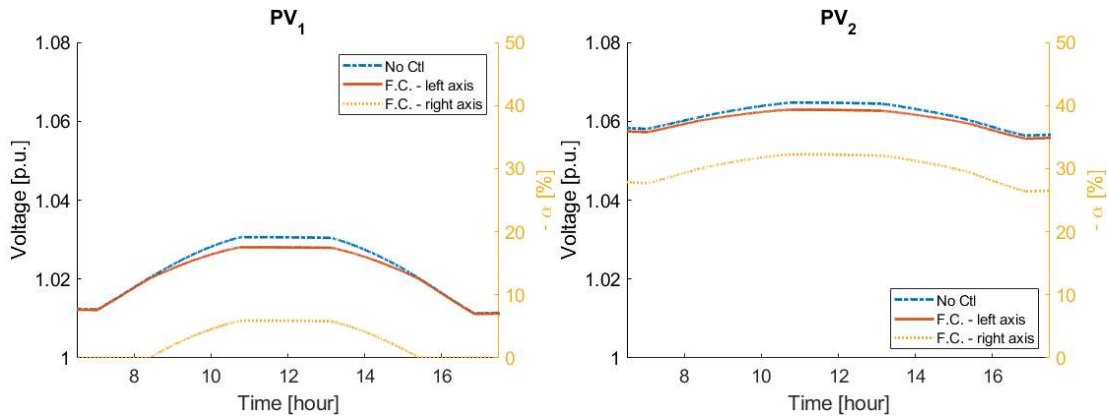

**Figure 3 Voltage profile (in per unit value) along a day for both PV systems under two different control conditions: no control (blue dashed line, No Ctl) and with the fixed curve (Volt-Var) as control (red line, F.C.). Reactive power fair utilization ratio profile ($\alpha$) for the fixed curve is included as well (dashed line, F.C.). Recall SI does not dispatch reactive power for no control case.**

### 3. Distributed optimization method

The proposed method takes advantage of reactive power dispatch and real power curtailment capacities of SIs (agents). The communication capabilities of these devices allow us to implement a distributed collaborative optimization to minimize voltage deviation at each agent's PCC with reactive power. Additionally, information exchange and its implementation on each SI yield to a unified voltage profile from a cooperative control perspective. On the other hand, real power curtailment on each agent is performed to have access to the full inverter capacity, if it is necessary.

As voltage deviation minimization is the goal of the method to be implemented, then the objective function can be defined as

$$F_v = \sum_m f_{v_m}, \quad \text{(eq. 3)}$$

$$f_{v_m} = \frac{1}{2}(1 - V_m)^2, \quad \text{(eq. 4)}$$

where $f_{v_m}$ is the objective function component associated with agent $m$ of the distributed network, and $V_m$ is the voltage at PCC of agent $m$. Maknouninejad & Qu (2014) proposes equation 3 can be minimized applying a distributed collaborative optimization using $\alpha_m$ as the control variable, and, in this way, $\alpha_m$ can be considered as an estimation of the optimal solution for that equation. At time $t_k$, the methodology establishes the agent $m$ maintains $\alpha_m(k)$ as its estimate of the solution. Then, it will calculate a new estimation for the next time step, $\alpha_m(k+1)$, using its own value from the previous time, and estimations performed by other units with whom there is communication, $\alpha_j(k)$ (with $j$ representing all agents that exchange information with agent $m$). The mathematical formulation of this relation is defined as follow,

$$\alpha_m(k+1) = d_{mm}\alpha_m(k) + \sum_{j \neq m} d_{mj}\alpha_j(k) - \beta_m g_m \quad \text{(eq. 5)}$$

$\beta_m > 0$ is the step size gain for agent $m$, and $g_m$ is the gradient of $f_{v_m}$ with respect to agent $m$'s state, $\alpha_m$, and is calculated as Maknouninejad & Qu (2014) as follows

$$g_m = -\bar{Q}_m(1 - V_m)\frac{V_m}{Q_m - V_m^2 B_{mm}}, \quad \text{(eq. 6)}$$

where $B_{mm}$ is the susceptance of node $m$ and it is defined as the sum of the imaginary parts of line conductances, corresponding to all lines connected to node $m$. This formulation for the gradient was derived using system power flow equations for real and reactive power.

On the other hand, $d_{mj}$ (and $d_{mm}$) are the communication coefficients, defined as

$$d_{mj} = \frac{w_{mj}s_{mj}}{\sum_l w_{ml}s_{ml}} \qquad j, l = 1..M \quad \text{(eq. 7)}$$

The coefficients $s_{mj}$ and $w_{mj}$ (in the same way that $s_{ml}$ and $w_{ml}$) belong to the communication topology matrix and the communication weight matrix, respectively, and they are associated with the physical connections existent on the distribution feeder. $M$ is the number of agents connected to the distribution grid, and $s_{ij}=1$ represents the information exchange existence between agents $i$ and $j$ (similarly, $s_{ij}=0$ means no communication between agents $i$ and $j$). From cooperative control theory, matrix $S$, which contains all $s_{ij}$ associated with a particular distribution grid, must contain at least one globally reachable node as a minimum requirement in order to ensure convergence of the represented distribution network (Xin et al., 2011). In other words, the proposed method is robust under scenarios of variations and limitations of the communication network. On the other hand, matrix $W$ stores $w_{ij} > 0$; each of them corresponding to the communication weight associated to the existing communication link between agent $i$ and $j$. This relation can be read as "how relevant the state of agent j in time step $t_k$ is for the state of agent $i$ in time step $t_{k+1}$, with respect to other agents connected to $i$ in time step $t_k$."

This work proposes changing the communication weights along the time (dynamic weights), and their values will be set based on their capacity to determine the influence of one agent into another. This will be utilized to allow agents with reactive power available helping in the voltage regulation of another agent, connected to a different bus of the distribution grid, which doesn't have enough reactive power capacity to do it independently. To implement this behavior, matrix $W_D$ is defined as follows

$$W_D = \begin{bmatrix} \bar{Q}_{1T}/\bar{Q}_1 & s_{12}\cdot\bar{Q}_1/\bar{Q}_2 & s_{13}\cdot\bar{Q}_1/\bar{Q}_3 & \ldots & \ldots & s_{1M}\cdot\bar{Q}_1/\bar{Q}_M \\ s_{21}\cdot\bar{Q}_2/\bar{Q}_1 & \bar{Q}_{2T}/\bar{Q}_2 & & & & s_{2M}\cdot\bar{Q}_2/\bar{Q}_M \\ s_{31}\cdot\bar{Q}_3/\bar{Q}_1 & & \ddots & & & s_{3M}\cdot\bar{Q}_3/\bar{Q}_M \\ \vdots & & & \ddots & & \vdots \\ \vdots & & & & \ddots & \vdots \\ s_{M1}\cdot\bar{Q}_M/\bar{Q}_1 & s_{M2}\cdot\bar{Q}_M/\bar{Q}_2 & s_{M3}\cdot\bar{Q}_M/\bar{Q}_3 & \ldots & \ldots & \bar{Q}_{MT}/\bar{Q}_M \end{bmatrix}$$

$\bar{Q}_{mT} = \sum_j s_{mj} \bar{Q}_j$ is the total reactive power available considering all SIs in communication with agent $m$. Each element outside the diagonal in the matrix $W_D$ can be described as the relative reactive power available of a specific agent with respect to each of its neighbors, where weight equal to zero when no communication exists between those agents. In general, $W_D$ will assign a big number to $w_{ij}$ if agent $i$ has several times more reactive power available than agent $j$. On the other hand, elements in the diagonal of the matrix $W_D$ correspond to the fraction of total reactive power available considering all agents that are communicated with a specific agent, with respect to its own reactive power available. In the same way, a big number in the diagonal of $W_D$ means that agent experiences reactive power scarcity.

The implementation of a distributed collaborative optimization on SIs can be interpreted as an adaptive Volt-Var curve that varies along the time based on the reactive power availability of each SI connected to a distribution grid. Then, each time step equation 5 will provide a new amount of reactive power that needs to be dispatch for an SI, based on its PCC voltage, maximum reactive power available, current reactive power dispatched, and susceptance. In the next section, to prove the effectiveness of the dynamic weights implemented on the mentioned distributed collaborative optimization, results combining the adaptive curve with different ways to define communication weights will be presented.

## 4. Simulation study and results

To show the effectiveness of the proposed method, proof-of-concept simulations are run on a modified IEEE 123 node test feeder, with two PV systems connected: one with an inverter size of 1500 kVA and 1800 kW of PV panels ($PV_1$) and the second one with a power rating capacity of 75 kVA coupled with 90kW of PV nominal capacity ($PV_2$). They are connected to nodes 250 and 450, respectively; both PVs are 3-phase systems (Figure 1). The nominal voltage for IEEE 123 node test feeder is 2.4 kV. The irradiance profile used as input corresponds to clear sky day and it is shown in Figure 4. The peak load is 3.49 MW of real power and 1.92 MVAr of reactive power and changes along the day according to the load profile of Figure 4. To facilitate result interpretations, regulators were removed from the distribution feeder, and transformer control functions were disabled as well. In addition, a 1000 kVAr capacitor was added to node 450 in order to increase the voltage deviation at that node. The simulation time step selected was 10 seconds, SIs update their output every 20 seconds, and the simulation is carried out for 11 hours period of the day from 6:30 and 17:30.

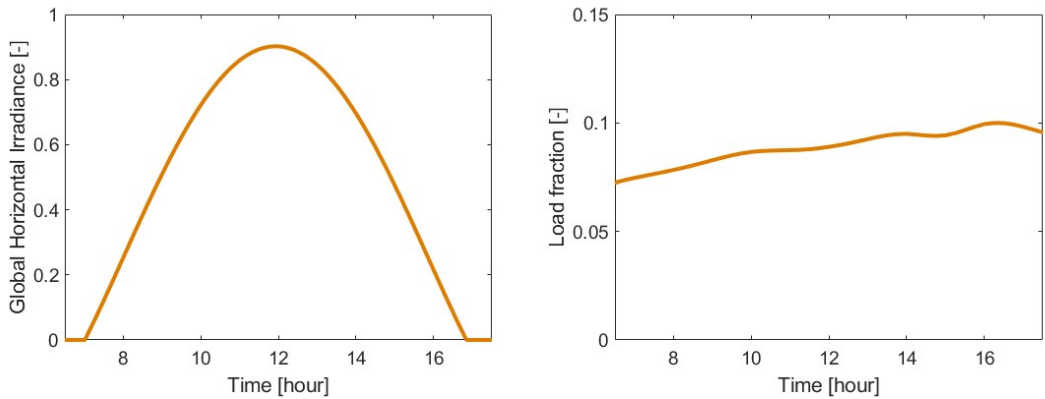

**Figure 4 Global horizontal irradiance (left) and load (right) profile utilized for simulations.**

In addition to one case where no reactive power is dispatched from any smart PV inverter, established as the no control case (No Ctl), four different control schemes implemented on each SI are simulated in this work. The first one uses fixed Volt-Var curve with parameters corresponding to the default settings of IEEE 1457 standard norm and it will be named fixed curve (F.C.). The other three control methods provide the reactive power generation based on the distributed optimization defined by equation 5 (adaptive curve, A.C.), but considering different communication topologies (no communication or communication) and methods to calculate communication weights (fixed or dynamic). One control method does not incorporate any information exchange between the SIs, and it corresponds to a full decentralized approach (adaptive curve with no communication, A.C.NoCm). The second control method incorporates communication between SIs but considers all communication weights fixed and equal to 1 (adaptive curve with fixed weights, A.C.F.W.). Finally, the last case implements the proposed method where communication exists between the agents with communication weights that changes along the time (adaptive curve with dynamic weights, A.C.D.W.). All methods presented modify parameters related to the AC side of the inverter, then there is no modification applied to the MPPT control utilized by the PV panels connected in the DC side. Table 1 shows the agent estimation update for each control method implemented.

Table 1 Agent estimation update formulation for each control method simulated. $f(V_m)$ represents the Volt-Var curve presented in Figure 2.

| Name | Abbreviation | Agent estimation update |
|---|---|---|
| No control | No Ctl | $\alpha_m(k+1) = 0$ |
| Fixed curve | F.C. | $\alpha_m(k+1) = \dfrac{f(V_m)}{\bar{Q}_m}$ |
| Adaptive curve with no communication | A.C. NoCm | $\alpha_m(k+1) = \alpha_m(k) - \beta_m g_m$ |
| Adaptive curve with fixed weights | A.C.F.W. | $\alpha_m(k+1) = \dfrac{1}{2}[\alpha_m(k) + \alpha_j(k)] - \beta_m g_m$ |
| Adaptive curve with dynamic weights | A.C.D.W. | $\alpha_m(k+1) = d^D{}_{mm}\alpha_m(k) + d^D{}_{mj}\alpha_j(k) - \beta_m g_m$, $d^D{}_{mj} = \dfrac{w^D{}_{mj}}{w^D{}_{m1} + w^D{}_{m2}}, W_D = [w^D{}_{ij}]$ |

Figure 5 shows the voltage profile of each PV system with each tested control method. Although all of them reduce voltage deviation for both SIs, with respect to the no control case, the reference value is reached only on $PV_1$ applying the A.C.NoCm method. The same method for $PV_2$ does not perform as well as on $PV_1$, and methods combining adaptive curve with communication provide better voltage regulation in $PV_2$. A.C.D.W. increases voltage deviation on $PV_1$ and decreases it on $PV_2$ with respect to A.C.F.W.

As each SI is undersized by 17%, flat real power generation during the time interval 10:45 – 13:08 in Figure 6 indicates inverter saturation. In addition, differences between each control method and no control profile in Figure 6 represents the amount of energy curtailed on each case. For $PV_1$, almost no curtailment is registered for fixed curve control method, while some real power fraction is reduced with other methods, and adaptive curve with dynamic weights provides the lowest real power generation. For $PV_2$, adaptive curve methods with no communication (A.C.NoCm) and dynamic weights (A.C.D.W.) curtail 100% for the entire period, while there is only partial curtailment for F.C. and A.C.F.W. methods.

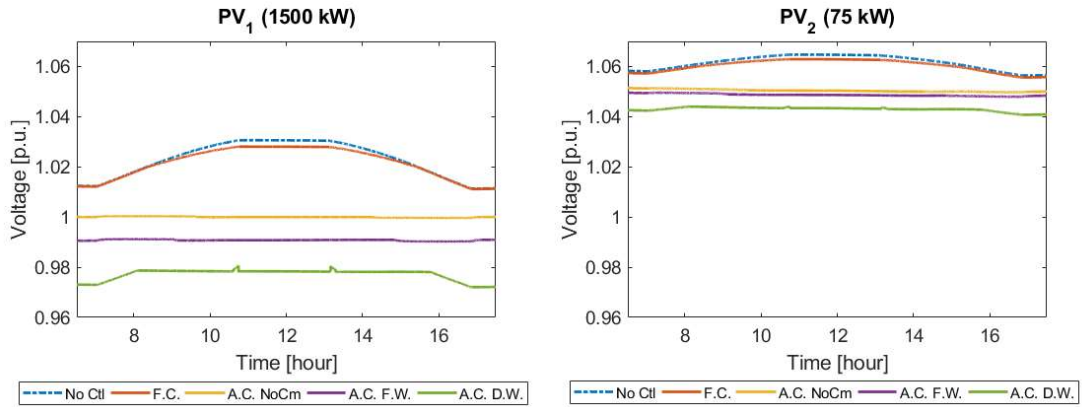

Figure 5 Voltage deviation (in per unit value) for $PV_1$ (left) and $PV_2$ (right) considering the no control case (No Ctl) and four different control methods: fixed control curve (F.C.), adaptive curve with no communication (A.C. NoCm), adaptive curve with fixed weights (A.C. F.W.), and adaptive curve with dynamic weights (A.C. D.W.).

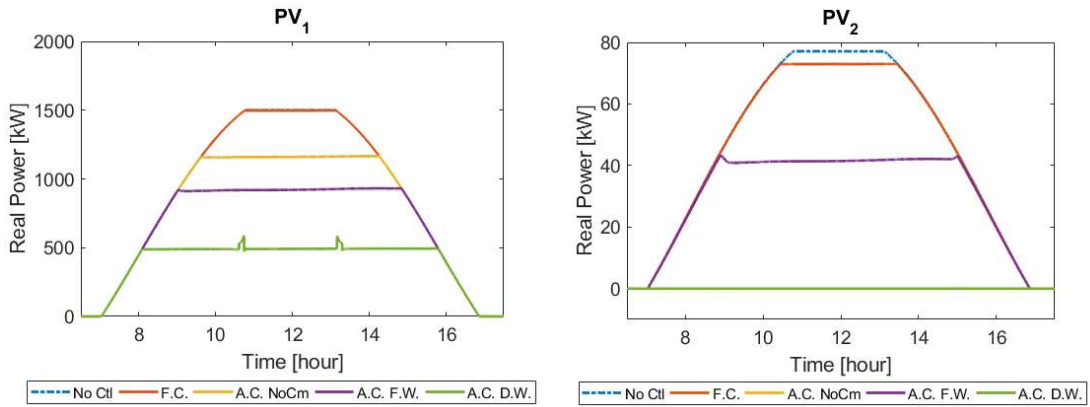

Figure 6 Real power generated for each PV system considering four the control methods plus no control case. NoCtl and F.C. real power results overlap in $PV_1$. A.C.NoCm and A.C.D.W. real power results overlap in $PV_2$.

Reactive power fair utilization ratio for each SI is presented in Figure 7. While no method utilizes full $PV_1$ inverter capacity to dispatch reactive power, A.C.NoCm and A.C.D.W. methods use full SI capacity to absorb reactive power in $PV_2$. This behavior indicates $PV_2$ reactive power scarcity to regulate its voltage. Under this scenario, methods with communication demand more reactive power absorption from $PV_1$ than methods-based on local measurements. In other words, the extra reactive power absorbed by $PV_1$ in methods with communication is dispatched to contribute to the voltage regulation in $PV_2$. Additionally, A.C.D.W. method requests a higher amount of reactive power absorption than A.C.F.W., which indicates dynamic weights are incorporating $PV_2$ reactive power scarcity to the solution and making $PV_1$ absorbs even more reactive power than the case of fixed weights.

Figure 8 shows a comparison between all methods considering objective function final result, for each control method applied, in two ways: adding $f_v$ for all agents connected to the distribution grid for each time step, and the final objective function result across the entire time period squared ($F_v^2$). Adaptive curve methods show an important improvement in voltage deviation across the day with respect to F.C. and no control methods. In particular, A.C.D.W. provides the lowest voltage deviation for the entire time period, where main differences occur during daylight. This difference comes from the additional reactive power absorbed by $PV_1$, which decreases voltage deviation in $PV_2$ with a small increment in the deviation of $PV_1$ voltage.

Results for average losses along the time period in active and reactive circuit elements are presented in Figure 9. All control methods implemented decreased the losses in comparison with No Ctl but the highest variation is lower than 20%. A.C.NoCm obtained the smallest losses among all methods.

Percentages of real power curtailed with each method on each SI with respect to the No Ctl case are presented in Figure 9 as well. Results presented on this figure show the same behave observed in Figure 6: all A.C. methods increased the amount of curtailment; and A.C.D.W. obtained the highest amount of energy curtailed on both SIs, which is consistent with the method that requested reactive power the most. In addition, $PV_2$ curtailed all the reactive power generated along the day for A.C.NoCm and A.C.D.W. methods. These results suggest future work should incorporate some curtailment minimization in equation 4 to improve the solution.

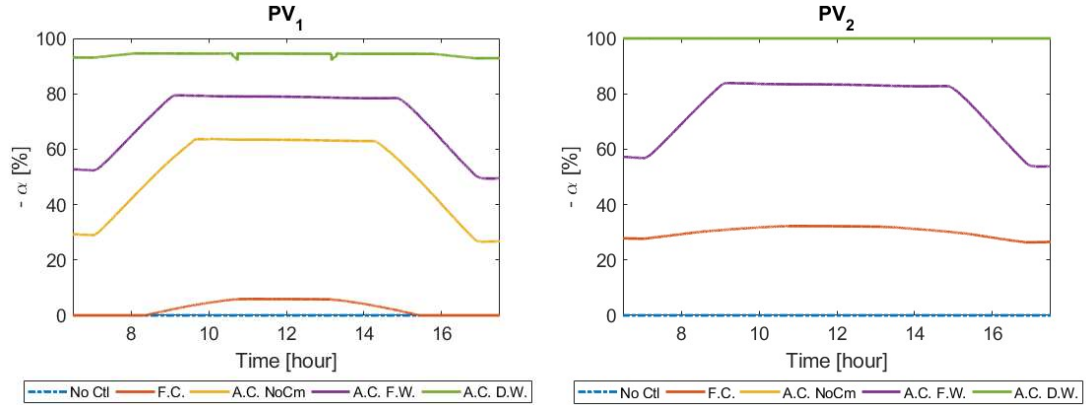

Figure 7 Reactive power fair utilization ratio ($\alpha$) dispatched from each SI for the no control case, and the four control methods tested.

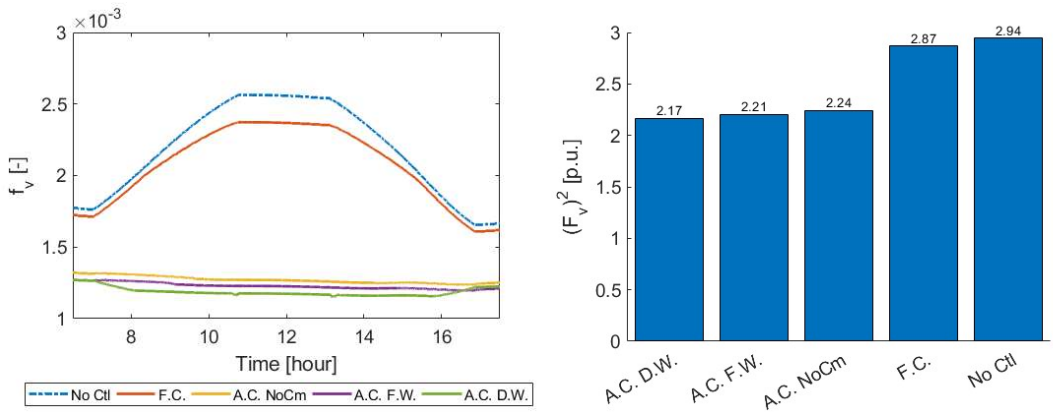

Figure 8 (Left) Sum of all objective function elements from each agent ($f_v$, equation 4), for each time step for all control methods plus no control case. (Right) the objective function results squared ($F_v^2$, equation 3), adding the entire time period for each method tested.

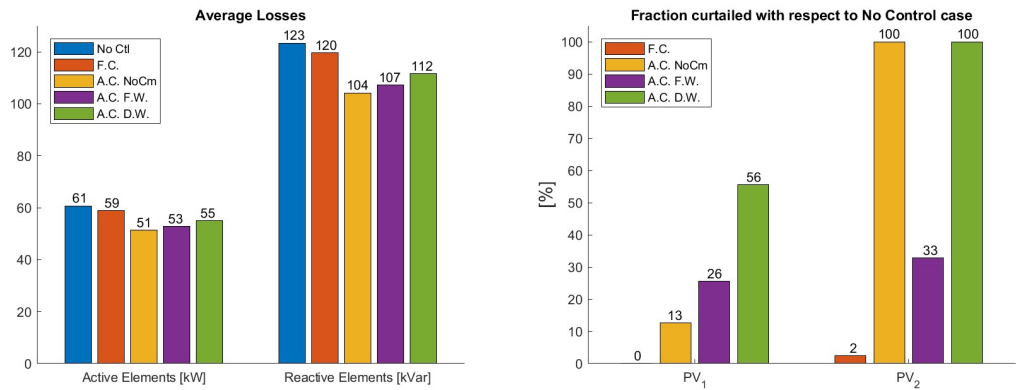

Figure 9 (Left) Average of the total losses for the active and reactive circuit elements for all control methods. (Right) Percentage of active power curtailed for each method on each SI with respect to the No Ctl case.

## 5. Conclusion

A distributed collaborative optimization technique with dynamic communication weights is proposed to regulate the voltage across a distribution power grid with high PV penetration. The main idea is to make the control structure adaptive to generation and load changes. Proof-of-concept simulations on a modified IEEE 123-node test feeder with two PV systems compare the effectiveness of the proposed approach to the existing techniques: fixed control curve, adaptive curve with no communication, and adaptive curve with fixed weights. Case studies show that the dynamic weight-based approach results in the lowest voltage deviation reducing losses with respect to No Ctl case, and without increasing losses significantly with respect to the other methods implemented. However, the main disadvantage of A.C.D.W. is the high amount of curtailment applied on each SI. This can be improved in future work incorporating a term for curtailment on the objective function. More realistic scenarios, with real feeder models and numerous PV systems, will also be studied. Irradiance profiles with high variability along the day will be incorporated in future work as well, to test oscillations and power quality impacts of the proposed method.

## 6. Acknowledgments

CC acknowledges financial support from CONICYT PFCHA/DOCTORADO BECAS CHILE/2017.